\documentstyle[art12,fleqn]{article}
\def\soc{{\rm C}_{60}}
\def\rug{{\rm C}_{70}}

\def\beeq{\begin{equation}}
\def\eneq{\end{equation}}
\def\beeqa{\begin{eqnarray}}
\def\eneqa{\end{eqnarray}}

\addtocounter{section}{-1}
\begin{document}

\begin{center}

{\large {\bf{Nonlinear optical response in higher fullerenes}}}

\mbox{}

{\rm Kikuo H{\sc arigaya}}\\
{\sl Physical Science Division, Electrotechnical Laboratory,\\
Umezono 1-1-4, Tsukuba 305, Japan}

\end{center}

\mbox{}

\noindent
{\bf Abstract}

Nonlinear optical properties of extracted higher fullerenes
-- C$_{70}$, C$_{76}$, C$_{78}$, and C$_{84}$ -- are 
theoretically investigated.  Magnitudes of off-resonant 
third-harmonic-generation are calculated by the intermediate 
exciton theory.  We find that optical nonlinearities of 
higher fullerenes are a few times larger than those of $\soc$.  
The magnitudes of nonlinearity tend to increase as the 
optical gap decreases in higher fullerenes.

\mbox{}


\newpage

It has been found that C$_{60}$ thin films show large optical 
nonlinearities.$^{\rm 1-4)}$  The observation is attractive
in view of scientific interests as well as technological 
applications.  The magnitude of the third-order nonlinear 
susceptibility, $\chi^{(3)}_{\rm THG} (\omega) = \chi^{(3)} 
(3\omega;\omega,\omega,\omega)$, for third harmonic generation 
(THG) is of the order of $10^{-12}$ esu to $10^{-11}$ esu.  
This large response is comparable to the values measured 
in polydiacetylenes.  The optical spectra of C$_{70}$$^{4)}$ and
higher fullerenes (C$_{76}$, C$_{78}$, C$_{84}$, etc.)$^{5,6)}$ 
have also been obtained. In order to explain the interesting 
experiments, several theoretical investigations$^{\rm 7-11)}$ 
have been performed.  We have applied a tight binding model$^{7)}$ 
to $\soc$, and have analyzed the nonlinear optical properties.  
Coulomb interaction effects on the absorption spectra and the 
optical nonlinearity have been also studied.$^{10)}$  We have 
found that the linear absorption spectra of $\soc$ and $\rug$ 
are well explained by the Frenkel exciton picture$^{11)}$ except 
for the charge transfer exciton feature around the excitation 
energy 2.8 eV of the $\soc$ solids.$^{12)}$  Coulomb interaction 
effects reduce the magnitude of the optical nonlinearity from 
that of the free electron calculation,$^{10)}$ and we have 
discussed that the local field enhancement might be effective 
in solids.

In the previous paper,$^{13)}$ we have investigated geometric 
effects on optical properties in higher fullerenes.  We have 
obtained the optical absorption spectra at a certain combination
of the pentagonal carbons, by using projected wavefunctions 
onto selected pentagons in order to calculate dipole moments.
The contributions from a part of the fullerene to the optical 
spectra have been extracted.  We have found that the optical 
excitations in the energy region smaller than about 4 eV have 
most of their amplitudes at the pentagonal carbons.  The 
oscillator strengths of absorption projected onto these carbons 
almost accord with those of the total absorption.  We have 
also found that the spectral shapes of the total absorption are 
mainly determined by the geometrical distributions of the 
pentagons in the fullerene structures.

The main purpose of this paper is to investigate nonlinear 
optical properties in higher fullerenes, further.  The 
calculation method of the THG has been used for $\soc$ in 
the previous work.$^{10)}$  Recently, a calculation of the 
THG in isomers of C$_{78}$ by a free electron model has 
been reported.$^{14)}$  However, the Coulomb interaction 
effects, whose importance we have found in C$_{60}$, 
have not been discussed for higher fullerenes.  The present 
paper gives rise to a new contribution to this subject.  We 
focus on the off-resonant THG in order to estimate possible
magnitudes of the nonlinearities of each isomer.  The 
Coulomb interaction strengths are also changed in a 
reasonable range, because realistic strengths are not well
known in higher fullerenes.

In the present paper, the carbon network of the fullerene 
surface is taken into account by the hopping integral $t$ 
between nearest neighbor sites.  The Coulomb interactions 
are taken into account by the parametrized Ohno potential, 
$W(r) = 1 / \sqrt{ (1/U)^2 + (r/r_0 V)^2 }$, between two 
electrons with distance $r$.  Here, $U$ is the interaction 
strength at the distance $r=0$, $V$ is the strength of the 
long range part, and $r_0$ the mean bond length.  The 
Coulomb interaction is treated by the restricted Hartree-Fock 
approximation and the intermediate treatment of excitons.$^{11)}$  
The THG is calculated by the sum-over-states method which
has been used in ref. 7.  In calculating the expectectation 
values of the dipole moment, the lattice coordinates contained
in the geometry package of higher fullerenes$^{15)}$ are used.  
We will discuss properties of the THG in seven kinds of the 
extracted fullerene isomers: C$_{70}$, C$_{76}$, C$_{78}$, 
and C$_{84}$.  From our experiences of investigation on 
optical properties,$^{10,11)}$ we can assume $V = U/2$ in 
the following.  The onsite Coulomb strength is varied within 
$0 \leq U \leq 4t$.

Figure 1 shows the absolute value of the off-resonant THG
$| \chi^{(3)}_{\rm THG} (0)|$ plotted against the Coulomb
interaction strength $U$.  The different plots indicate
differenet kinds of isomers.  The four isomers -- C$_{70}$, 
$D_2$-C$_{76}$, $D_3$-C$_{78}$, and one kind of 
$C_{2v}$-C$_{78}$ [$C_{2v}$ by Kikuchi's notation (ref. 16)] --
exhibit similar magnitudes of optical nonlinearities.
On the other hand, the other three isomers -- one kind of 
$C_{2v}$-C$_{78}$ [$C_{2v}^{'}$ by Kikuchi's notation (ref. 16)], 
$D_{2d}$-C$_{84}$, and $D_2$-C$_{84}$ -- show larger optical 
nonlinearities than those of the former isomers.  This is 
mainly due to the smaller energy gap of the latter isomers, 
even though the negative correlation between the THG and 
the energy gap is not so complete through all the isomers.
The decrease of THG from the free electron model ($U = 0$)
to the case with $U=4t$ is by the factor about 0.1 for all 
the isomers.  The similar fact has been found in the 
calculation of $\soc$.$^{10)}$  This would be a general 
property in various kinds of higher fullerenes.  The overall 
magnitudes of the THG are around 10$^{-12}$ esu for most of 
the Coulomb interactions considered.

In Fig. 2, the relations between the absolute value of the 
off-resonant THG $| \chi^{(3)}_{\rm THG} (0)|$ and the energy 
gap are shown for three Coulomb interaction strengths: $U = 0t$, 
$2t$, and $4t$.  Here, the energy gap is defined as the optical 
excitation energy of the lowest dipole allowed state.  This is 
called the optical gap, in other words.  For each Coulomb 
interaction, the seven plots cluster in a bunch.  When the 
energy gap becomes larger, the THG tends to decrease.  However, 
the correlation between the THG and the energy gap is far
from that of a smooth function.  The correlation is merely a 
kind of tendency.  Therefore, the decrease of the energy gap 
in higher fullerenes is one of origins of the larger optical 
nonlinearities of the systems.  The actual magnitudes would 
also be influenced by detailed electronic structures of isomers.

In the calculations of $\soc$, the magnitudes of the THG at 
the energy zero are about $1 \times 10^{12}$ esu in the free 
electron model (ref. 7), and about $1 \times 10^{-13}$ esu 
for $U = 4t$ and $V = 2t$ (ref. 10).   In the present 
calculations of higher fullerenes, the magnitudes are a few 
times larger than those of $\soc$.  Thus, this paper predicts 
that nonlinear optical responses in higher fullerenes are 
generally larger than in $\soc$.  In the previous paper,$^{10)}$ 
we have discussed that the local field correction factor 
is of the order 10 in $\soc$ solids.  As the distance between the 
surfaces of neighboring fullerene molecules in C$_{70}$ and 
C$_{76}$ solids is nearly the same as in C$_{60}$ solids, 
we expect that local field enhancement in thin films of higher 
fullerenes is of the similar magnitudes as in $\soc$ systems.

In summary, we have investigated nonlinear optical properties
of higher fullerenes.  Theoretical off-resonant THG has been
calculated by using the intermediate exciton theory.  We have 
found the optical nonlinearities of higher fullerenes which
are larger than in $\soc$.  The magnitudes of THG tend to 
increase as the optical gap decreases in higher fullerenes.

\pagebreak

\noindent
{\bf References}

\mbox{}

\noindent
1) J. S. Meth, H. Vanherzeele, and Y. Wang: Chem. Phys. Lett. 
{\bf 197} (1992) 26.\\
2) Z. H. Kafafi. J. R. Lindle, R. G. S. Pong, F. J. Bartoli, 
L. J. Lingg, and J. Milliken: Chem. Phys. Lett. {\bf 188} (1992) 492.\\
3) F. Kajzar, C. Taliani, R. Danieli, S. Rossini, and R. Zamboni:
Chem. Phys. Lett. {\bf 217} (1994) 418.\\
4) B. C. Hess, D. V. Bowersox, S. H. Mardirosian, and
L. D. Unterberger: Chem. Phys. Lett. {\bf 248} (1996) 141.\\
5) R. Ettl, I. Chao, F. Diederich, and R. L. Whetten: Nature 
{\bf 353} (1991) 149.\\
6) K. Kikuchi, N. Nakahara, T. Wakabayashi, M. Honda, H. Matsumiya,
T. Moriwaki, S. Suzuki, H. Shiromaru, K. Saito, K. Yamauchi,
I. Ikemoto, and Y. Achiba: Chem. Phys. Lett. {\bf 188} (1992) 177.\\
7) K. Harigaya and S. Abe: Jpn. J. Appl. Phys. {\bf 31} (1992) L887.\\
8) E. Westin and A. Ros\'{e}n: Int. J. Mod. Phys. B {\bf 23-24}
(1992) 3893.\\
9) Z. Shuai and J. L. Br\'{e}das: Phys. Rev. B {\bf 46} (1992) 16135.\\
10) K. Harigaya and S. Abe: J. Limun. {\bf 60\&61} (1994) 380.\\
11) K. Harigaya and S. Abe: Phys. Rev. B {\bf 49} (1994) 16746.\\
12) S. L. Ren, Y. Wang, A. M. Rao, E. McRae, J. M. Holden, 
T. Hager, K. A. Wang, W. T. Lee, H. F. Ni, J. Selegue 
and P. C. Eklund, Appl. Phys. Lett. {\bf 59} (1991) 2678.\\
13) K. Harigaya and S. Abe: J. Phys.: Condens. Matter
{\bf 8} (1996) 8057.\\
14) X. Wan, J. Dong, and D. Y. Xing: J. Phys. B {\bf 30} 
(1997) in March issue.\\
15) M. Yoshida and E. \={O}sawa:
The Japan Chemistry Program Exchange, Program No. 74.\\
16) K. Kikuchi, N. Nakahara, T. Wakabayashi, S. Suzuki,
H. Shiromaru, Y. Miyake, K. Saito, I. Ikemoto, M. Kainosho,
and Y. Achiba: Nature {\bf 357} (1992) 142.\\

\pagebreak

\begin{flushleft}
{\bf Figure Captions}
\end{flushleft}

\mbox{}

\noindent
Fig. 1.  The absolute value of the off-resonant THG
$| \chi^{(3)}_{\rm THG} (0)|$ plotted against the Coulomb
interaction strength $U$.  The closed and open squares
are for C$_{70}$ and $D_2$-C$_{76}$.  The closed circles
are for $D_3$-C$_{78}$, and the open and crossed circles 
are for two kinds of $C_{2v}$-C$_{78}$.  The closed and
open triangles are for $D_{2d}$-C$_{84}$ and $D_2$-C$_{84}$,
respectively.

\mbox{}

\noindent
Fig. 2.  The absolute value of the off-resonant THG
$| \chi^{(3)}_{\rm THG} (0)|$ in seven isomers of
higher fullerenes, plotted against the energy
gap (shown in units of $t$).  The squares, circles, and
triangles are for $U = 0t$, $2t$, and $4t$, respectively.
The left axis is in the logarithmic scale.

\end{document}